\documentclass[aps,prl,superscriptaddress,twocolumn,floatfix]{revtex4}

\usepackage{amsmath}
\usepackage{amssymb}
\usepackage{bm,color}
\usepackage{graphicx,epsfig,color}

\begin{document}

\title{Response to the Comment on ``Excitons in Molecular Aggregates with L\'evy Disorder: Anomalous Localization and Exchange Broadening of Optical Spectra"}

\maketitle

In our Letter~\cite{Eisfeld10}, we predicted exchange broadening and a blue shift of the absorption spectrum with increasing disorder strength $\sigma$ for heavy-tailed disorder with a stability index $\alpha < 1$.
We explained these findings by deriving 'conventional' scaling laws for the width of the absorption spectrum and the delocalization length.
In addition we predicted new features in the density of states and the absorption spectrum, as well as non-universality of the localization length distribution, which we attributed  to the appearance of 'outliers' in the disorder, i.e. sites with energies outside of the bare exciton band.

The main points of
the Comment~\cite{Werpachowska12} are that the
outliers introduced by heavy tails in the disorder distribution (i)
do not lead to deviations from the conventional scaling law for the
half width at half maximum (HWHM) of the absorption spectrum and
(ii) do not lead to non-universality of the distribution of
localization lengths. We show
below that the findings reported in our Letter~\cite{Eisfeld10}
are correct and that the wrong conclusions of the Comment~\cite{Werpachowska12} arise from focusing on small $\sigma$
values.

The conventional scaling law for the HWHM ($\equiv$~FWHM/2) in the
general case of symmetric $\alpha$-stable L{\'e}vy distributions
reads 
\begin{equation}
{\rm HWHM}\sim J(\sigma/J)^{2\alpha/(1+\alpha)}.
\end{equation}
 Its derivation was given in our Letter~\cite{Eisfeld10} already,
generalizing from the special cases of Gaussian ($\alpha=2)$ and
Lorentzian ($\alpha=1)$ disorder \cite{Malyshev93,Vlaming09} to arbitrary values of $\alpha$. In the Comment~\cite{Werpachowska12}, this derivation is repeated without any additional insights (last paragraph of the Comment~\cite{Werpachowska12}). In
the derivation, segmentation due to outliers (defined by us as
sites with deviations from the average energy larger than $2 J$,
acting as barriers) as well as finite size effects are not taken
into account. For $\sigma$ values of interest for many
molecular systems ($\sigma \lessapprox J$), deviations from the
conventional HWHM scaling due to outliers have not been seen for
$\alpha=1$ or $2$. However, for $\alpha\approx 0.5$ and smaller clear
deviations are visible in this $\sigma$ regime. This is due to the
fact that for the same $\sigma$, outliers become more abundant for
heavier tails (smaller $\alpha$). By equating $N^*$ and
$\bar{N}_{\rm seg}$ from Eqs.~(3) and (4) of our Letter~\cite{Eisfeld10}, one can estimate the value of $\sigma$
above which segmentation may be expected to play a significant role
(indicated as the shaded region in Fig.~1a).

\begin{figure}[tbp]
\vspace{-0.5cm}
\includegraphics[width=0.95\columnwidth]{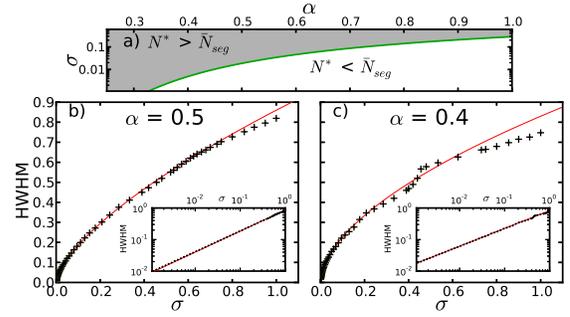}
\caption{(a): ``phase diagram'' showing for which
($\alpha$,\,$\sigma$) combinations segmentation is expected to become
visible.
 (b,c): HWHM of the absorption peak as a function of $\sigma$ for
(b) $\alpha=0.5$  and (c) $\alpha=0.4$. Data points obtained from
numerical simulations ($N=200$, 10$^7$ realizations), red curve
representing a fit corresponding to the conventional localization
mechanism (note the deviations from the fit). }
\label{fig0.4a}
\end{figure}

To demonstrate the deviations from the conventional scaling at
smaller $\alpha$ values, we plot in Fig.~1 the HWHM as a function of
$\sigma$ for $\alpha=0.5$ (b) and $\alpha=0.4$ (c). 
The red lines are power-law fits imposing the conventional scaling
law.
The case $\alpha=0.5$, panel (b), was also given in our Letter~\cite{Eisfeld10} but now has many more data points (symbols), also compared to the Comment~\cite{Werpachowska12}.
For $\sigma \lesssim 0.4 J$ the conventional power
law fit works well, agreeing with the observation in Ref.~\cite{Eisfeld10} and repeated in the Comment; for larger $\sigma$, the
conventional scaling clearly breaks down. Moreover, the data points
show kinks at which the HWHM suddenly jumps. This is not noise,
but may be unambiguously attributed to structure in the high-energy
wing of the absorption spectrum which grows in for increasing
$\sigma$ as a consequence of segmentation~\cite{Eisfeld10}. While
both effects  are visible already at $\alpha=0.5$, they become more
pronounced for smaller $\alpha$ (panel c). This new data also shows
that these features cannot be adequately represented by a power law.
Finally, we mention that the straight line in Fig.~2 of
our Letter~\cite{Eisfeld10} indeed is meaningless without the
information how the monomer HWHM scales with $\sigma$; this has no
effect on the conclusion of that figure, namely the discovery of
exchange broadening.

\begin{figure}[b]
\includegraphics[width=0.9\columnwidth]{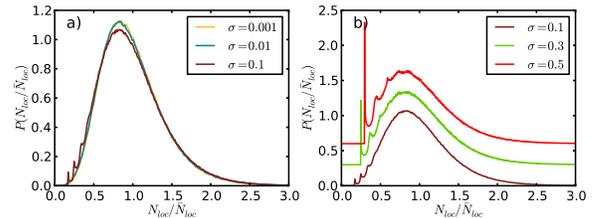}
\caption{Distribution of normalized localization lengths for small
disorder values (a) and intermediate disorder values (b). In (b) the
curves have different vertical offsets to better distinguish their
differences. Notably, additional structure appears at the low
localization length side, corresponding to segmentation-limited wave
functions. $N=200$, $\alpha=0.5$. } \label{fig:Loc}
\end{figure}
%

Next we demonstrate the non-universality of the localization length
distribution due to the segmentation by outliers. While the data
used for this purpose in our Letter~\cite{Eisfeld10} (Fig.~3 and the
$\sigma$-dependence of $\delta N_{loc}/{\bar N}_{loc}$) were
correct, they were not a good choice, especially not because they
were discussed in the context of Ref.~\cite{Vlaming09}, where
the energy interval in which states are considered was scaled with
the width of the absorption band. In Fig.~2, we show that also for
an energy interval which is ``properly'' scaled, the
non-universality becomes visible. This figure displays for
$\alpha=0.5$ the distribution of normalized (to the average)
localization lengths, obtained using the same scaling of the width
of the  energy interval as in the Comment~\cite{Werpachowska12} (cf.~Ref.~
\cite{Vlaming09}). In Fig.~2(a) small $\sigma$ values are
considered ($\sigma \le 0.1 J$, as in the Comment~\cite{Werpachowska12}). As expected from Fig.~1(a)
segmentation is not relevant then and the normalized distributions
are nearly identical for all $\sigma \le 0.1 J$. Yet, already for
$\sigma=0.1 J$ deviations become clearly visible in the appearance
of several small sharp peaks at small localization length. Upon
increasing $\sigma$ (Fig.~2(b)), these deviations become more and
more pronounced and the non-universal character of the distribution
is clear
\footnote{When taking a broader (``correctly'' scaled) energy range, the segmentation and the non-universality becomes even more pronounced.}. 
By considering the wavefunctions, the origin of
these peaks can be traced back to the segmentation mechanism. We
note that also for the fixed energy interval used in
our Letter~\cite{Eisfeld10}, these signatures of segmentation and
non-universality can be found.

In conclusion, we have shown that the claims put forth in the Comment
 are not generally valid, but are
instead an artifact of focusing on a regime where segmentation
effects are expected to be small. In the Comment it is stated that conventional
scaling may break down at high $\sigma$ values, only because
$N_{loc}$ approaches unity. This is a trivial effect and does not explain the additional structure observed in the localization length distributions. As we have
demonstrated, the breakdown found by us is more subtle and really is
intimately related to outliers.

\vspace{0.2cm}

\noindent
A.~Eisfeld$^1$, S.M.~Vlaming$^{2,3}$, S.~M\"obius$^1$, V.A.~Malyshev$^2$ and J.~Knoester$^2$

{\footnotesize
$^1$Max Planck Institute for Physics of Complex Systems,
N\"othnitzer Strasse 38, D-01187 Dresden, Germany

$^2$Centre for Theoretical Physics and Zernike Institute
for Advanced Materials, University of Groningen, Nijenborgh 4, 9747
AG Groningen, The Netherlands

$^3$Center for Excitonics and Department of Chemistry, Massachusetts Institute of Technology,
77 Massachusetts Avenue, Cambridge, Massachusetts 02139, USA
}

\end{document}